# Intrinsic Regularization in a Lorentz invariant non-orthogonal Euclidean Space

Carmen Tornow
*Institute of Planetary Research, German Aerospace Center, Rutherfordstr. 2, D-12489 Berlin*
E-mail: carmen.tornow@dlr.de

**Abstract**

It is shown that the Lorentz transformations can be derived for a non-orthogonal Euclidean space. In this geometry one finds the same relations of special relativity as the ones known from the orthogonal Minkowski space. In order to illustrate the advantage of a non-orthogonal Euclidean metric the two-point Green's function at x = 0 for a self-interacting scalar field is calculated. In contrast to the Minkowski space the one loop mass correction derived from this function gives a convergent result due to an intrinsic regularization parameter called effective dimension. This parameter is an entropy related measure for the information loss caused by quantum fluctuations of the metric at energies higher than the Planckian limit.

## I. Introduction

Ultra-violet divergences in quantum field theory (QFT) can occur for momentum integrals $\int d^4k$ over the Fourier transformed Green's functions. These expressions are related to the Feynman diagrams with closed loops symbolizing the creation and annihilation of virtual particles which originate from the vacuum. As *off mass-shell* particles they have a momentum range $k_\mu \in [0,\infty)$. A typical example is the one-loop diagram of the fermion and boson propagator representing the self-energy of the corresponding particle. Its contribution diverges for $k_\mu \to \infty$ due to an asymptotic behavior proportional to $\int d^4k/k^2$. A further interpretation of this integral requires a regularization and renormalization procedure (Collins, 1995). The first step separates the finite and infinite parts of the diagram. The second one introduces redefined parameters, e.g., for the particle mass or coupling constant which absorb the isolated divergent terms. As a result, the renormalized Lagrangian resembles the original one if the QFT is renormalizable. However, the redefined parameters are infinite for free and finite for interacting quantum fields.

All renormalization methods have a common origin, they are caused by the incompleteness of QFT as a low energy approximation of a more complex theory. The regularization scheme illustrates this incompleteness by introducing an additional parameter, e.g. the upper limit of the momentum *K* in cut-off regularization, a fictive particle having an upper limit of mass *M* in Pauli-Villars regularization, the grid-size $\chi$ since the space-time is approximated by a 4-D hyper-cube in lattice regularization, and a reduction of space-time dimensions from 4 to 4 - $\kappa$ in dimensional regularization. The implementation of a lowest distance, a highest momentum or the manipulation of the space-time dimensionality suggest that the usual geometry of QFT - the orthogonal Minkowski space (OMS) - may not be quite appropriate.

## II. The Lorentz transformations in the NES

Based on these considerations the question arises whether there is an alternative geometry that supports Lorentz invariance. The preferred metric should be influenceable by the momentum of the particle in a way that for an increasing energy the number of dimensions relevant to the integration decreases. Thus, a non-orthogonal Euclidean space (NES) with energy dependent angles between its axes could result. In this NES an effective dimensionality varying between one and four would arise as regularization parameter. Due to its effect one should get a finite solution for the above mentioned one-loop diagram symbolizing the interaction between a real and a virtual particle whereas the simple $\varphi^4$ – QFT is chosen as test case.

At first, a method is presented which shows the existence of a Lorentz invariant NES. In order to validate this method it is used to derive the Lorentz transformations (LT) for the OMS (see

Appendix 1). Two inertial systems $\Sigma$ and $\Sigma'$ moving relative to each other with the velocity **v** can be arranged in a way, that the velocity vector is, say, $\mathbf{v}=V\mathbf{e}_{x1}$, i.e., $x \equiv x_1 \neq x'_1 \equiv x'$, $x_2 = x'_2$ and $x_3 = x'_3$. The origin of $\Sigma'$ coincides with the mass centre of the moving particle. For the selected two-parameter approach of the coordinate transformations between $\Sigma$ and $\Sigma'$ one has to apply the two following conditions in order to fix these parameters:

1. In order to maintain an isotropic space it is postulated that the time-related coordinate $\tau = ct$ transforms similar to the spatial coordinate x.
2. The length of a vector should be the same in $\Sigma$ and $\Sigma'$, i.e., transformation independent.

The first condition is true for both OMS and NES. Hence, one starts with the Equations (1.3) and (1.4) given in Appendix 1. Replacing the parameters $\alpha$, $\beta$ by $\omega$, $\sigma$ it follows

$$x' = \omega'(x - \sigma\tau) \quad \text{and} \quad \tau' = \omega'(\tau - \sigma x) \quad \text{and} \quad \omega\omega' = (1-\sigma^2)^{-1}. \tag{1}$$

Due to the special arrangement $\mathbf{v}=V\mathbf{e}_x$ the moving inertial system $\Sigma'$ belongs to the 2-D NES denoted by $_n\mathbf{E}^2(\rho)$ whereas the free parameter $\rho$ is used to introduce a metric **g** according to

$$\mathbf{g} = \frac{1}{\lambda}\begin{pmatrix} 1 & \rho \\ \rho & 1 \end{pmatrix} \quad \text{and} \quad \lambda = (1-\rho^2)^{1/2}. \tag{2}$$

The determinant of **g** is $|\mathbf{g}| = 1$ and the geometrical meaning of the parameter $\rho$ results from

$$\rho = \cos(\angle(\mathbf{e}_x, \mathbf{e}_t)) \tag{3}$$

whereas $\mathbf{e}_x$ and $\mathbf{e}_t$ are the unit vectors of the x and $\tau$ axes. In Figure 1 three selected examples of $\Sigma'$ are presented. The angles correspond to three different values for $\rho$: 90°, 60°, and 30°.

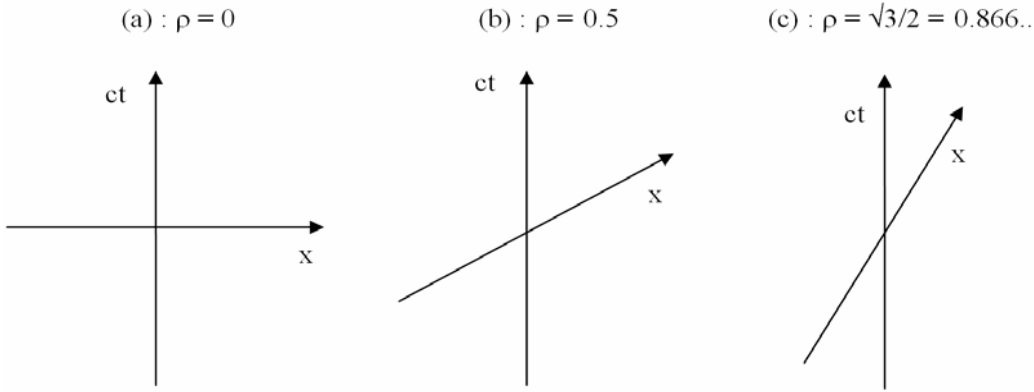

*Figure 1: Influence of different $\rho$ values in the NES shown in three examples of $\Sigma'$.*

Equivalently to Equation (1.5) in Appendix 1, the length conservation for the vectors $(\tau, x)^T$ and $(\tau', x')^T$ in the NES leads to

$$\tau^2 + x^2 = (\tau'^2 + x'^2 + 2\rho\tau'x')/\lambda. \tag{4}$$

Inserting Equation (1) into (4) and considering that the term $\tau'x'$ must vanish one finds

$$4\sigma = 2\rho(1+\sigma^2) \quad \text{and} \quad \omega'^2(1+\sigma^2 - 2\rho\sigma) = \lambda \tag{5}$$

from which the dependency between $\rho$, $\lambda$ and $\sigma$ results according to

$$\rho = 2\sigma/(1+\sigma^2) \quad \text{and} \quad \lambda = |1-\sigma^2|/(1+\sigma^2) \quad \text{and} \quad \omega = \omega' = |1-\sigma^2|^{-1/2}. \tag{6}$$

Therefore, from Equation (1) it follows

$$x' = (x - \sigma\tau)/\sqrt{|1-\sigma^2|} \quad \text{and} \quad \tau' = (\tau - \sigma x)/\sqrt{|1-\sigma^2|}. \tag{7}$$

A comparison of the Equation (7) with (1.7) in Appendix 1 shows the strong similarity of these expressions. To finish the confirmation that the transformations in Equation (7) are the LT, it is necessary to find a relation between $\rho$, $\sigma$ on the one hand and the ratio $\beta = V/c$ on the other.

Hereunto, one introduces the 2-velocity $\mathbf{u} = (u_0, u_1)^T$ attached to the moving particle. For the resting observer one gets $\mathbf{u^0} = (c, 0)^T$. According to Equation (7) the components of $\mathbf{u}$ are

$$\mathbf{u} = \omega(c, -\sigma c)^T. \tag{8}$$

A further relation for the two velocity components $u_a$, with $a = 0,1$ can be derived from

$$u_a = \frac{dx_a}{dt_e} = \frac{dx_a}{const \cdot dt} \Rightarrow \mathbf{u} = \frac{1}{const}(c, -V) \quad \text{with} \quad \frac{dx_1}{dt} = -V, \tag{9}$$

whereas $dt_e$ is the differential "Eigenzeit" of the particle and $\mathbf{x} = (x_0, x_1)^T = (\tau, x)^T$. The light speed $c$ results from $dx_0 \equiv d\tau = cdt$. If one compares $\mathbf{u}$ in (8) and (9) one finds

$$dt_e = const \cdot dt = \omega^{-1} dt = \sqrt{|1-\sigma^2|} \, dt \quad \text{with} \quad \sigma = V/c = \beta. \tag{10}$$

Since $\sigma \equiv \beta$, it follows that the transformations in Equation (7) are the LT. The 2-momentum $\mathbf{p}$ is given by $\mathbf{p} = (p_0, p_1)^T$ whereas $p_0 = E/c$ and E is the total energy of the moving particle. One calculates E using the "Eigenzeit" $dt_e$

$$E = cm_0 u_0 = cm_0 \omega \frac{dx_0}{dt} = \frac{m_0 c^2}{\sqrt{|1-\sigma^2|}}. \tag{11}$$

Equation (11) is the energy-mass relation well-known from the special relativity in the OMS. Using the Equations (6) and (11) the metric parameter $\lambda$ can be expressed as

$$\lambda = m_0^2 c^4 / (2E^2 - m_0^2 c^4). \tag{12}$$

The energy of the moving particle influences the metric of the NES attached to it. For a nearly resting particle with $E \sim m_0 c^2$, the Euclidean space is almost orthogonal, i.e., $\lambda \sim 1$. Due to the Equation (10) one realizes that the angle $\rho$ between the axes in $\Sigma'$ depends on the velocity V of the moving particle relative to the resting observer. Since V is a measure for the particle energy the angle is energy dependent, just as required in the beginning of this section.

## III. The generalization of the NES

Next it is shown that the LT in the NES form a group consisting of three elements. Moreover, it will be allowed that both, the observer and the particle, are attached to moving systems. For this case, the addition theorem of velocities can be derived. The general form of Equation (4) can be written as

$$x_\alpha g^{\alpha\beta} x_\beta = x'_\mu g'^{\mu\nu} x'_\nu \quad \text{with} \quad x'_\mu = N_\mu^{\ \alpha} x_\alpha \tag{13}$$

whereas $\mathbf{x} = (x_0, x_1, x_2, x_3)^T$ and $x_0 \equiv \tau$. Based on this expression, it results a set of non-linear relations between the matrix components of $\mathbf{g}$, $\mathbf{g'}$ and $\mathbf{N}$

$$g^{\alpha\beta} = N_\mu^{\ \alpha} g'^{\mu\nu} N_\nu^{\ \beta} \quad \text{and} \quad N_\alpha^{\ \mu} N_\nu^{\ \alpha} = \delta_\nu^\mu. \tag{14}$$

The transformation matrix $N_\alpha^{\ \mu}$ depends on the velocity vector $\mathbf{v} = V_j \mathbf{e_j}$ of the observer system $\Sigma$ and the vector $\mathbf{v'} = V'_j \mathbf{e_j}$ is related to the particle system $\Sigma'$. Note, that $j = 1, 2, 3$ is arbitrary but fixed. The metric in $\Sigma$ depends on $\rho_j$ and is given by

$$g^{00} = g^{jj} = \lambda_j^{-1}, \quad g^{kk} = 1, \quad g^{j0} = g^{0j} = \rho_j \lambda_j^{-1} \quad \text{where} \quad \lambda_j = \sqrt{1-\rho_j^2} \quad \text{and} \quad k \neq j. \tag{15}$$

All the other elements of $\mathbf{g}$ are zero. An equivalent expression is valid for $\Sigma'$ using $\rho'_j$ and $\lambda'_j$ instead of $\rho_j$ and $\lambda_j$. For each index j a matrix $\mathbf{N_J}$ with $J \equiv j$ exists that is an element of the Lorentz group. Accordingly, the Lorentz group of the NES has three elements analogue to the OMS. Since the index J is equal to j it can be suppressed in the following, i.e., $\mathbf{N_J} \equiv \mathbf{N}$ and

$$N_0^{\ 0} = N_j^{\ j} = \omega_j, \quad N_k^{\ k} = 1, \quad N_0^{\ j} = N_j^{\ 0} = -\sigma_j \omega_j \quad \text{where} \quad \omega_j = |1-\sigma_j^2|^{-1/2} \quad \text{and} \quad k \neq j. \tag{16}$$

All the other elements of **N** are zero. The relation between $\sigma_j$ and $\rho_j$, $\rho'_j$ are derived from the Equation (14) which can be written explicitly as

$$(N_0^0)^2 + 2\rho'_j N_0^0 N_j^0 + (N_j^0)^2 = \frac{\sqrt{1-\rho'^2_j}}{\sqrt{1-\rho_j^2}} \quad \text{and} \quad \rho'_j(N_0^0)^2 + 2N_0^0 N_j^0 + \rho'_j(N_j^0)^2 = \rho_j \frac{\sqrt{1-\rho'^2_j}}{\sqrt{1-\rho_j^2}}. \qquad (17)$$

Due to the absolute value for $\omega_j$ in Equation (16) the following two solutions $\sigma_j$ and $\hat{\sigma}_j$ result

$$\sigma_j = (\rho'_j \lambda_j - \rho_j \lambda'_j)/(\lambda_j + \lambda'_j) \quad \text{or} \quad \hat{\sigma}_j = (\rho'_j \lambda_j + \rho_j \lambda'_j)/(\lambda_j - \lambda'_j). \qquad (18)$$

whereas $\sigma_j < 1$ and $\hat{\sigma}_j > 1$. Since we have shown in the previous section that in the NES the LT are applicable to relate $\Sigma$ to $\Sigma'$ it is reasonable to suggest that the two solutions correspond to the time- and space-like vectors known from the OMS. This suggestion can be underlined by the addition theorems of velocities derived next. From Equation (6) the approach

$$\sigma_j^0 = \rho_j/(1+\lambda_j) \quad \text{and} \quad \hat{\sigma}_j^0 = (1+\lambda_j)/\rho_j \qquad (19)$$

follows. The index 0 indicates that Equation (19) is valid for a moving and a resting system in contrast to Equation (18) which is applicable for two moving systems. Analogue expressions result for $\sigma'^0_j$ and $\hat{\sigma}'^0_j$ which depend on $\rho'_j$ and $\lambda'_j$. Inserting Equations (19) into (18) it follows

$$\sigma_j = (\sigma'^0_j - \sigma^0_j)/(1 - \sigma'^0_j \sigma^0_j) \quad \text{and} \quad \hat{\sigma}_j = (1 - \hat{\sigma}'^0_j \hat{\sigma}^0_j)/(\hat{\sigma}'^0_j - \hat{\sigma}^0_j). \qquad (20)$$

According to Equation (10) the transformation parameters $\sigma^0_j$ and $\sigma'^0_j$ are equal to $\beta$ and $\beta'$. Thus, the first relation in Equation (20) is the addition theorem of velocities lower than $c$ and the second relation is the addition theorem of velocities higher than $c$. Even for $\hat{\sigma}^0_j > 1$ the metric parameter $\rho_j$ is bounded by $\rho_j < 1$ (Figure 2). This behavior ensures that the metric in Equation (14) is positive definite, i.e., the length of a vector **r** always fulfils $\|\mathbf{r}\| \geq 0$. One gets

$$\|\mathbf{r}_\pm\|^2 = (r_0^2 + r_j^2 \pm 2|\rho_j r_0 r_j|)/\lambda_j^2 \geq 0. \qquad (21)$$

The two plots in Figure 2 show the curves of $\rho_j$ and $\lambda_j$ versus the parameters $\sigma^0_j = \rho_j/(1+\lambda_j)$ and $\hat{\sigma}^0_j = (1+\lambda_j)/\rho_j$. Note the increase of $\rho_j$ at the left and the decrease of $\rho_j$ at the right plot.

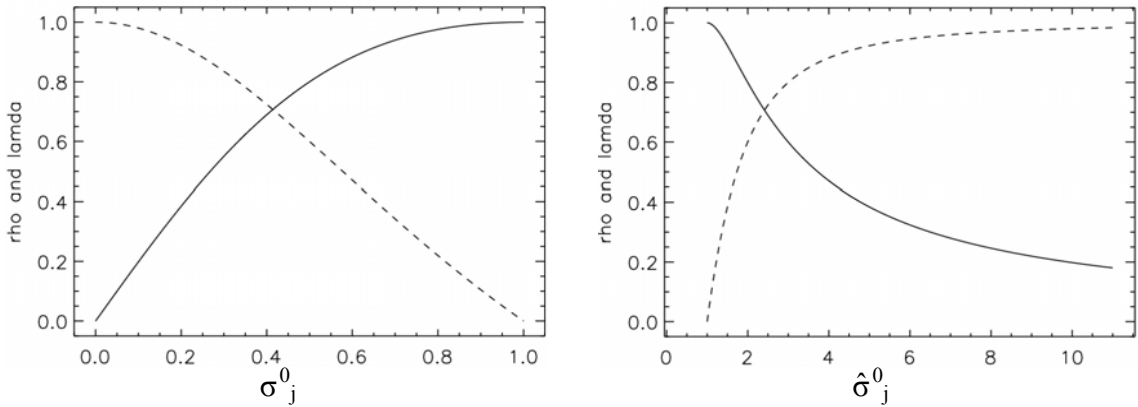

*Figure 2: Influence of the velocity ratio $\sigma^0_j$ and $\hat{\sigma}^0_j$ on the metric parameters $\rho_j$ (solid line) and $\lambda_j$ (dashed line) in the two different ranges with $V_j < c$ or $V_j > c$.*

Finally, one can show that the differential "Eigenzeit" in Equation (10) is directly calculated using Equation (15). Starting with

$$ds^2 = g^{\mu\nu} dx_\mu dx_\nu = (dx_0^2 + dx_j^2 + 2\rho_j dx_0 dx_j)/\lambda_j = c^2 dt_e^2 \quad \text{with} \quad dx_k^2 = 0, \quad k \neq j \qquad (22)$$

one gets a relation which agrees with the one in the OMS

$$dt_e^2 = dt^2 [1 - (\sigma^0_j)^2] \quad \text{with} \quad \sigma^0_j = \beta_j. \qquad (23)$$

The benefit of the geometry established in the NES $_n\mathbf{E}^4(\vec{\rho})$ is demonstrated in the next section.

## IV. The interacting scalar particle in the NES

Usually, a non-orthogonal space can be transformed to an orthogonal one if its metric is well-defined. However, in the NES one gets $\rho_j \to 1$ and $\lambda_j \to 0$ if $V_j \to c$, i.e., if the particle energy increases up to infinity for a rest mass $m_0 > 0$ (see Equation (12)). Hence, the 4 components of a vector, e.g., momentum, are weighted differently with a growing disparity. For all energies higher than the Planck energy quantum fluctuations of the metric occur which overwhelm the low weighted momentum components. To quantify the corresponding information loss, one introduces an effective dimension *q(E)* into the NES. Since *q(E)* acts as an energy dependent regularization parameter it allows a finite mass correction for the interacting scalar field. At first, one considers the 4-momentum vector **p** of a free particle and calculates its length giving

$$g^{\tau\nu} p_\tau p_\nu = m_0^2 g^{\tau\nu} u_\tau u_\nu = m_0^2 c^2. \tag{24}$$

The energy-momentum or *on mass-shell* expression is known from special relativity. Particles are real if they act according to Equation (24). Using the correspondence between momentum and spatial derivation, i.e., $p_\mu \equiv -i\hbar\, \partial/\partial x_\mu$, one finds the Klein-Gordon operator $\aleph$

$$\aleph\, \varphi(\mathbf{x}) = (\partial^\alpha \partial_\alpha + 1)\varphi(\mathbf{x}), \tag{25}$$

whereas $\varphi(\mathbf{x})$ describes a scalar particle, e.g. a Higgs boson. Introducing the inverse Compton length $l = m_0 \cdot c/\hbar$ the field $\varphi(\mathbf{x})$ is scaled as $\varphi \to \varphi/l$ with $[\varphi^2] = [\hbar]$ and $x_\mu \to l\, x_\mu$ with $[x_\mu] = 1$. For a self-interacting particle the action integral is given by

$$A[\varphi, J] = \int d^4 x \sqrt{-|g|}\left[-\varphi(\mathbf{x})\left(\partial^\alpha \partial_\alpha + 1\right)\varphi(\mathbf{x})/2 - V[\varphi] + J(\mathbf{x})\varphi(\mathbf{x})\right] \tag{26}$$

whereas $|g|=1$ (see explanations for Equation (2.1) in Appendix 2). $V[\varphi] = w/4! \cdot \varphi^4$ denotes the potential term with $[w] = [\hbar^{-1}]$. The Green's function of a free particle with $V[\varphi]=0$ is given by

$$D(\mathbf{x}) = \frac{1}{(2\pi)^4} \int d^4 k\, \frac{\exp(-i\mathbf{x}\mathbf{k})}{\mathbf{k}^2 - 1 + i\varepsilon} \quad \Leftrightarrow \quad \aleph D(\mathbf{x}) = -\delta(\mathbf{x}) \tag{27}$$

whereas $i\varepsilon$ is a regularization term for the denominator if $\mathbf{k}^2 = g^{\mu\nu} k_\mu k_\nu = 1$. The momentum scales as $k_\nu \to k_\nu/l$ with $[k_\nu] = 1$. In Appendix 2 it is shown that the Green's function $D(\mathbf{x}=0)$ is related to the first order mass correction (see Equation (2.5)). One gets the expression

$$\mu^* = \sqrt{1 + w\hbar\, \Re(D(0))/2} \quad \text{with} \quad D(0) = \frac{1}{(2\pi)^4} \int \frac{d^4 k}{\mathbf{k}^2 - 1 + i\varepsilon} \tag{28}$$

caused by the interaction of the real, scalar with the vacuum related, virtual particle (see the Equations (2.2) and (2.2a) in Appendix 2). Since the integral only depends on $\mathbf{k}^2$ one gets

$$D(0) = \frac{2}{(4\pi)^2 \Gamma[2]} \int \frac{dk\, k^3}{k^2 - 1 + i\varepsilon}. \tag{29}$$

The integration of Equation (29) comprehends the momentum range from null to infinity. Since the virtual particle is an *off mass-shell* particle and does not obey Equation (24) it can approach energies up to infinity. In this high momentum range, the equivalent relative velocity between the real scalar particle and the virtual one reaches *c*. Then, the resulting coordinate frame $\Sigma'$ in the NES adopts angles $\rho_j \sim 0°$. Its spatial axes $x_j$ carry low weights and momentum components corresponding to these axes are very small. The relevant intervals in Equation (29) extent

1. from $k = 0$ to $k = E_p/m_0 c^2$, and
2. from $k = E_p/m_0 c^2$ to $k \to \infty$ (Planckian range).

They differ from each other because in the Planckian range quantum fluctuations of the metric occur. These fluctuations can be larger than the low weighted momentum components. Due to these irresolvable uncertainties the LT cannot be performed anymore and the non-orthogonality of the Euclidean space turns into an intrinsic feature. The information loss for the particle momentum due to these uncertainties in the Planckian range have to be quantified. A suitable measure for an information loss caused by microscopic effects could be the entropy. It is used here to introduce an effective dimension $q(E)$ for the NES according to

$$q(E) = \exp\left[-\sum_{\alpha=0}^{3} \bar{e}_\alpha(E)\ln(\bar{e}_\alpha(E))\right] \quad \text{and} \quad \sum_{\alpha=0}^{3} \bar{e}_\alpha(E) = 1. \quad (30)$$

Equation (30) is based on the v. Neumann entropy (Balian, 2003) and derived in Appendix 3. The symbols $\{e_\alpha\}_{\alpha=0,3}$ denote the normalized eigenvalues of the metric, **g**, in Equation (15). According to the Equations (12) the metric parameter $\lambda_j$ and $\rho_j$ depend on the energy of the particle. This dependency also occurs for the eigenvalues given by

$$\bar{e}_{0,1}(E) = (1 \pm \rho_j)/[2(1+\lambda_j)] \quad \text{and} \quad \bar{e}_{2,3}(E) = \lambda_j/[2(1+\lambda_j)] \quad (31)$$

Due to Equation (30) the effective dimension is energy dependent as well. If $\rho_j = 0$, i.e., $\lambda_j = 1$, the eigenvalues are $\bar{e}_\alpha = ¼$ for all $\alpha$ and $q(E) = 4$. The metric is completely orthogonal. On the other hand, it is fully non-orthogonal with $q(E) \approx 1$ if $\rho_j \to 1$ so that $\bar{e}_0 \approx 1$ and $\bar{e}_\alpha \approx 0$ for $\alpha > 0$. Nevertheless, it is not the complete non-orthogonality which forbids the application of the LT but the occurrence of metrical fluctuations in the Planckian range that cover the low-weighted momentum components. Inserting $q(E)$ into the Equation (29) one gets

$$D(0) = \int dk \frac{2k^{q(E)-1}(k^2-1+i\varepsilon)^{-1}}{(4\pi)^{q(E)/2}\Gamma[q(E)/2]} \quad (32)$$

Now, the question arises, how this integral can be calculated. Figure 3 presents the dependency between effective dimension $q(E)$ and Planck energy ratio $E/E_P$. One sees, that for a particle with a low mass, i.e., $m_0 \ll E_P/c^2$ the dimension $q(E) \approx 1$ is reached long before the Planckian limit $E_P$ is met (left plot of Figure 3), i.e., long before metrical quantum fluctuations become important. Hence, LT can be applied even if $q(E) \approx 1$. For a heavy particle with $m_0 \sim E_P/c^2$ a different situation occurs since $q(E)$ changes from 4 to 1 at energy values near the Planckian limit (right plot of Figure 3). The quantum fluctuations of the metric have already started for $q(E) > 1$ and the LT as well as any other transformation can not fulfill Equation (13). A field theory describing the dynamics of this particle would require quantum gravity. Nonetheless, the self-interacting particle with the Lagrangian in Equation (26) should be fairly light with a mass comparable to the one of the Higgs boson, i.e., $m_0 \ll E_P/c^2$.

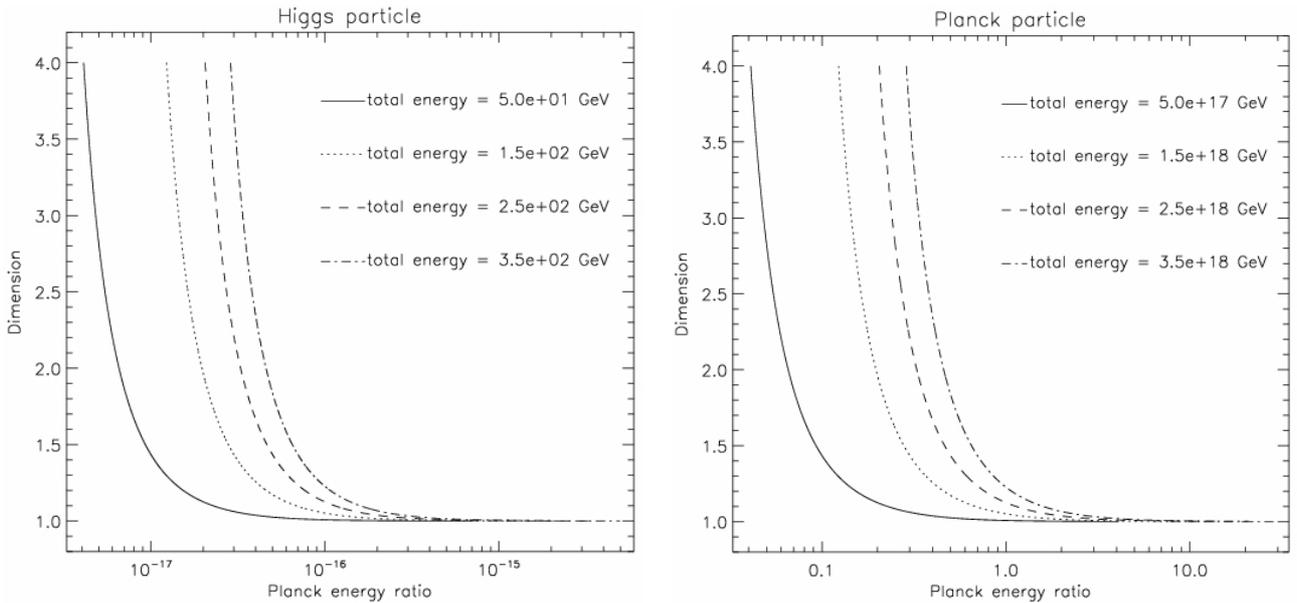

*Figure 3:* Dimension $q(E)$ versus the Planck energy ratio $E/E_P$ for a Higgs particle with different rest masses and a "Planck particle" with masses close to $E_P/c^2$.

In this case (E << $E_P$) the physical space is modeled by a 4-D orthogonal Euclidean space due to the applicability of the LT. If E ~ $E_P$ the LT are not allowed anymore but $q(E)$ has already reached $q(E) = 1$. As a result, one gets a jump function for Higgs-like particles given by

$$q(E) = 4 \quad \text{if} \quad E < E_P \quad \text{and} \quad q(E) = 1 \quad \text{if} \quad E \geq E_P \tag{33}$$

Due to the described behavior of $q(E)$ the integral in Equation (32) needs to be split into two parts. Since $q(E) = 1$ the second part extending over the Planckian range can be simplified. One realizes that the effective dimension acts as a regularization parameter since it influences the integration result in the high energy range. However, in contrast to the regularization procedures described in the Introduction the impact of $q(E)$ leads to a convergent result for the momentum integral in Equation (32). Therefore, no further renormalization is required. Using Equation (33) one gets the following expression for the Green's function D(0)

$$D(0) = \frac{1}{8\pi^2} \int_0^{k^*} \frac{dk\, k^3}{k^2 - 1 + i\varepsilon} + \frac{1}{\pi} \int_{k^*}^{\infty} \frac{dk}{k^2 - 1 + i\varepsilon}. \tag{34}$$

Because of the scaling one inserts for $k^*$ the ratio of the Planck to the rest energy of the Higgs particle, i.e., $k^* = E_P/m_0c^2$, whereas $m_0c^2 \equiv m_{Higgs}c^2 \leq 190$ GeV. Both terms are finite and give

$$D(0) = \frac{1}{16\pi^2}\left(k^{*2} + \ln(k^{*2} - 1) - i\pi\right) + \frac{1}{\pi}\left(\arctan(k^*) + i\pi/2\right) \approx \frac{k^{*2}}{16\pi^2} + 0.48i \tag{35}$$

whereas the small imaginary term can be absorbed by the ε-term in the denominator (compare Equation (2.5) in Appendix 2). Since the Planck energy is given by $E_P \approx 1.2 \cdot 10^{19}$ GeV one gets $k^* \leq 9.4 \cdot 10^{16}$. Then, the mass correction in Equation (28) leads to the following result

$$\mu^* = \sqrt{1 + w\Theta} \quad \text{with} \quad \Theta \approx \hbar \cdot \Re[D(0)]/2 \leq 2.95 \cdot 10^{-3} \text{Js}, \tag{36}$$

giving $w << \Theta^{-1} \approx 340$ Js$^{-1}$ due to the approximations made for the calculation of $\mu^*$ (Appendix 2). The lifetime of the scalar particle amounts $\tau_L \approx 0.34\,(w \cdot \text{Js})^{-1/2}$ ns (Equation (2.5) and below in Appendix 2). If the masses are clearly smaller than the Higgs mass, e.g., the electron mass $m_0c^2 \equiv m_ec^2 \approx 5.1 \cdot 10^{-4}$ GeV or the proton mass $m_0c^2 \equiv m_pc^2 \approx 0.94$ GeV, one gets a value of $\Theta \approx 1.8 \cdot 10^8$ Js or $\Theta \approx 55$ Js, respectively. That means, $w$ has to be arbitrary small or Equation (36) leads to inconsistent results. Consequently, the theory of a self-interacting scalar field as described in Equation (26) gives reasonable results for fairly heavy scalar particles. For light particles the ad-hoc introduced potential term V[φ] may be not useful.

**V. Conclusion**

It is shown that the LT applied in the NES lead to the known expressions of special relativity. Since the Lagrangian of each field theory is a scalar term one can derive the equations of motion and accordingly the Green's function in an orthogonal Euclidean space. As long as the energy is small enough to neglect the quantum fluctuations of the metric the derived results for the orthogonal geometry are equivalent to those attached to the non-orthogonal one since both metrics are related to each other by the LT (see Equations (13) and (14)).

If the momentum components of a virtual particle enter the Planckian limit the LT cannot be applied anymore. Then, non-orthogonality of the Euclidean space becomes an intrinsic feature, and the contributions of the momentum components linked to the low-weighted axes cannot be separated from the quantum fluctuations of the metric. Due to the corresponding information loss, the utilization of an effective dimension $q(E)$ in the NES (see Equation (30)) makes sense since $q(E)$ is smaller than the geometrical dimension. The calculations in Section 4 show that $q(E)$ acts as a intrinsic regularization parameter. In contrast to the regularization parameters in the OMS $q(E)$ is not introduced ad-hoc. It seems that the NES considers high energy processes more appropriately than the OMS. The non-orthogonal space in combination with its effective dimension allows to determine the first-order mass correction as well as the lifetime for a self-interacting scalar particle defined in Equation (26). The results in the Equations (35) and (36) are more conceivable if the scalar particle has a fairly high *free particle mass* $m_0$.

However, to develop a complete quantum field theory in an Euclidean space it is necessary to include particles having a spin larger than 0, such as vector bosons or spinors. For instance, only a few changes are required to get the electromagnetic tensor for an Euclidean geometry. Therefore, as a next step, it will be useful to develop the field equations for a spin-½ particle combined with a Maxwell field as gauge boson.

**Acknowledgement**

The author would like to thank Uwe Motschmann from the Institute for Theoretical Physics at the Technical University Braunschweig for many fruitful discussions.

**Appendix 1 – the LT in the OMS**

Using equivalent notations as in the NES one starts with the following two-parameter approach

$$x' = \alpha'(x - \beta\tau), \quad x = \alpha(x' + \beta\tau') \quad \text{with} \quad \beta = V/c. \tag{1.1}$$

This approach fulfils the requirement that the Galilei transformations are valid for $|v| \ll c$. So, Equation (1.1) ensures that $x = \beta\tau$ for the *spatial* origin of $\Sigma'$ expressed in $\Sigma$ and $x' = -\beta\tau'$ for the *spatial* origin of $\Sigma$ expressed in $\Sigma'$. Inserting x' into the relation for x it results for $\tau'$

$$\tau' = [(1 - \alpha\alpha')/\alpha\beta]x + \alpha'\tau. \tag{1.2}$$

The requirement that the time and space coordinates are transformed in a similar way leads to

$$x' = \alpha'(x - \beta\tau) \quad \leftrightarrow \quad \tau' = \alpha'(\tau - \beta x). \tag{1.3}$$

A comparison between the Equations (1.2) and (1.3) gives

$$(1 - \alpha\alpha')/\alpha\beta = -\alpha' \quad \text{or} \quad \alpha\alpha' = (1 - \beta^2)^{-1}. \tag{1.4}$$

The second constraint arises from the conservation of the length of a vector in $\Sigma$ and $\Sigma'$. Thus

$$\tau^2 - x^2 = \tau'^2 - x'^2. \tag{1.5}$$

If one inserts the expressions for x' and $\tau'$ from Equation (1.3) into the r.h.s. of Equation (1.5) and using Equation (1.4) for $\alpha\alpha'$ the following relation can be derived

$$\alpha = \alpha' = (1 - \beta^2)^{-1/2}. \tag{1.6}$$

Consequently, one gets the well-known LT

$$x' = (x - \beta\tau)/\sqrt{1 - \beta^2} \quad \text{and} \quad \tau' = (\tau - \beta x)/\sqrt{1 - \beta^2}. \tag{1.7}$$

For the spatial origin of $\Sigma'$ the time-related $\tau'$ fulfills $\tau' = \tau(1 - \beta^2)^{1/2}$ and equivalently for $\Sigma$ and $\tau$.

**Appendix 2 – the relation between Green's function and mass correction**

The connected 2-point Green's function for a self-interacting particle can be calculated from

$$\Delta_c(\mathbf{x}_1, \mathbf{x}_2) = \frac{\hbar \delta^2 \ln(Z[J])}{\delta J(\mathbf{x}_1) \delta J(\mathbf{x}_2)}\bigg|_{J=0} \quad \text{and} \quad Z[J] = \frac{\int D\varphi \exp\left[-\frac{1}{\hbar} A[\varphi, J]\right]}{\int D\varphi \exp\left[-\frac{1}{\hbar} A[\varphi, 0]\right]}. \tag{2.1}$$

whereas $A[\varphi, J]$ is the action integral in Equation (26). In contrast to the OMS, in the NES the expression $(-|\mathbf{g}|)^{1/2} = i$. So, the action integral has to be multiplied by *i* for the NES in order to substitute $|\mathbf{g}|$. Accordingly, the path integral gives exp(-A/ℏ) in contrast to the OMS where one gets exp(iA/ℏ) since there $(-|\mathbf{g}|)^{1/2} = 1$. The mass correction for the scalar particle is expressed

using the Greens function. The corresponding calculations are described in good books on path integrals in QFT and can be adapted to the NES. One obtains

$$\Delta_c(\mathbf{x}_1, \mathbf{x}_2) = D(\mathbf{x}_1 - \mathbf{x}_2) + \frac{\hbar w}{2} D(0) \int d^4\mathbf{z}\, D(\mathbf{z} - \mathbf{x}_1) D(\mathbf{z} - \mathbf{x}_2). \tag{2.2}$$

in a first order calculation. It corresponds to the so called dressed propagator represented by

$$\overline{x_1 \quad x_2} = \overline{x_1 \quad x_2} + \underset{x_1 \quad x_2}{\bigcirc}. \tag{2.2a}$$

The Fourier transform of Equation (2.2) gives

$$\Gamma_c(\mathbf{p}, \mathbf{q} = -\mathbf{p}) = (\mathbf{p}^2 - 1 + i\varepsilon)^{-1} + \frac{\hbar w}{2}(\mathbf{p}^2 - 1 + i\varepsilon)^{-2} \int \frac{d^4\mathbf{k}}{(2\pi)^4} (\mathbf{k}^2 - 1 + i\varepsilon)^{-1}. \tag{2.3}$$

For a sufficient weak interaction $w\hbar \cdot D(0)/2 \ll 1$ it can be written in the following form

$$\Gamma_c(\mathbf{q}, -\mathbf{q}) = \frac{1}{\mathbf{q}^2 - [1 + w\hbar/2 \cdot \Re(D(0))] + i[\varepsilon - w\hbar/2 \cdot \Im(D(0))]} \tag{2.4}$$

In Equation (2.4) the denominator contains the terms $1 + w\hbar \cdot \Re(D(0))/2$ and $w\hbar \cdot \Im(D(0))/2$, which can be related to the corrected mass $\mu^{*2}$ and to the inverse lifetime $\tau^*_L{}^{-2}$ of the self-interacting particle. Note that both parameters are scaled; thus one gets

$$\mu^* = \sqrt{1 + w\hbar \Re(D(0))/2} \quad \text{and} \quad \tau^*_L = \sqrt{2/w\hbar \Im(D(0))}. \tag{2.5}$$

Since $\Im(D(0)) = 0.48$ (see Equation (35)) the un-scaled lifetime amounts to $\tau_L = \hbar/m_0 c^2 \cdot \tau^*_L$ which gives to $\tau_L \approx 3.4 \cdot 10^{-10} (w/Js)^{-1/2}$ s or $\tau_L \approx 0.34 (w/Js)^{-1/2}$ ns.

**Appendix 3 – the effective dimension**

The effective dimension is a measure for the information content that can be attached to the measuring results of an experiment. If the NES has four dimensions, it is possible to get the maximum information content for all four components of a vector. The variability of, say, the energetic state of a particle would tend to zero if the measuring result of its energy-momentum vector is kept fixed. Otherwise, if the NES has only about one effective dimension one could get an independent information for only one component. Keeping this measuring result fixed no strong reduction of the internal variability of the system can be expected.

The entropy is appropriate to quantify the information content of measurements regarding the internal variability of a system. Consequently, one can try to use the entropy to determine the effective dimension. Von Neumann (1932) has proposed the entropy of a quantum mechanical system in the following way

$$S_N(\eta) = -k_B \operatorname{Trace}[\eta \log(\eta)] \quad \text{with} \quad \eta_{kl} = |\theta_k\rangle\langle\theta_l| \tag{3.1}$$

whereas $\eta$ is the density matrix of an ensemble of quantum mechanical states $\theta_k$. However, a consideration of the information content of measurements also leads to the definition given by Shannon. Indeed, the relation to Shannon's entropy (1948) becomes apparent if the quantum states are assumed to be orthogonal. Then, one has a density matrix $\eta_{kl} = \omega_k \delta_{kl}$ and it follows

$$S_{Sh}(\eta) = -k_B \Sigma_k \omega_k \log(\omega_k) \quad \text{with} \quad 1 = \Sigma \omega_k. \tag{3.2}$$

The eigenvalues $\omega_k$ if normalized in the given way can be linked to the probabilities of the states $\theta_k$. For an equal distribution of N states in the ensemble, one gets

$$S_{Sh}(\eta) = k_B \log(N). \tag{3.3}$$

Equation (3.3) can be related to the entropy expression of Boltzmann (1866)

$$S_B(\eta) = k_B \log(\Omega) \tag{3.4}$$

whereas Ω is the phase space volume associated with a given macro-state $\Psi_L$ corresponding to the integral of the Liouville volume element. In the discrete case the macro-state $\Psi_L$ is given by L microstates $\{\psi_k\}_{k=1,L}$. Each $\psi_k$ is a $q_k$-fold degenerate state. One gets for Ω

$$\Omega = M!/q_1! \cdot q_2! \cdots q_L! \quad \text{with} \quad \Sigma_k q_k = M \tag{3.5}$$

which for large numbers M and $q_k$ leads to

$$S_B/M = k_B \log(\Omega)/M = -k_B \Sigma_k (q_k/M) \log(q_k/M). \tag{3.6}$$

using the expression $\log(x!) = x \log(x) - x$ for both M and $q_k$. Since the ratio $q_k/M$ is equal to the probability $\omega_k$ in Equation (3.2) all entropy definitions are equivalent

$$S_{Sh} = S_B/M = S_N. \tag{3.7}$$

The study of fractal sets (Jizba & Arimitsu, 2001) have given an insight into the relationship between entropy and dimension whereas the relation used here is modified a little bit. Due to the quantum fluctuations of the metric the space-time continuum seems to loose its continuous character on its smallest scale and appears as a (multifractal?) set from the standpoint of the measuring results. Using Equation (3.7) one derives for the effective dimension $N \equiv q_N$ from

$$\log(q_N) \equiv \log(\Omega^{1/M}) = -\text{Trace}[\eta \log(\eta)] \tag{3.8}$$

Let's write $\eta = UEU^T$, whereas E is a diagonal matrix with the normalized eigenvalues $\bar{e}_i$ of η in the main diagonal axis. Since the von Neumann entropy fulfils $S(\eta) = S(U\eta U^T)$ for a real density matrix η, Equation (3.8) gives

$$q_N = \exp[-\Sigma_\alpha \bar{e}_\alpha \ln(\bar{e}_\alpha)] \quad \text{with} \quad 1 = \Sigma_\alpha \bar{e}_\alpha \tag{3.9}$$

whereas the effective number of states $q_N$ is related to the phase space volume by $q_N = \Omega^{1/M}$. If the metric **g** can be viewed as some density matrix for fluctuating states (Equation (4.9) in Appendix 4) the effective number of macroscopic, i.e., measurable states $q_N$ in Equation (3.9) is identical to the effective dimension $q(E)$ introduced in Equation (30).

**Appendix 4 – blurred LT due to metrical fluctuations**

Let's assume that the metric $g^{\mu\nu}$ is influenced by quantum fluctuations which cause a non-continuous space. Then, the matrix **N** should fulfill the following transformation property

$$X'_\mu = N_\mu^\alpha X_\alpha + c_\mu + \varsigma_\mu \quad \text{with} \quad g^{\mu\nu} \langle \varsigma_\mu \varsigma_\nu \rangle \Rightarrow \text{Min}. \tag{4.1}$$

whereas Σ and Σ' differ in their relative velocity $v_j$, j=1,2,3 in a same way as for the LT. Thus one can call Equation (4.1) blurred LT with the additional requirement that the elements of **N** minimize the transformation error $\langle \varsigma_\mu \varsigma^\mu \rangle$ produced by the metrical fluctuations. It is not the aim of this Appendix to treat $g^{\mu\nu}$ as a quantum field operator. However, it is conceivable that in a first order approximation $X_\mu$ and $X'_\mu$ appear as blurred space-time coordinates. In other words, the metric fluctuations $g^{\mu\nu} = G^{\mu\nu} + \xi^{\mu\nu}$ are redefined as coordinate fluctuations $x_\mu$

$$X_\mu = G_{\mu\nu} X^\nu + \xi_{\mu\nu} X^\nu = \langle X_\mu \rangle + x_\mu. \tag{4.2}$$

whereas $\langle X'_\mu \rangle$ and $\langle X_\mu \rangle$ are the macroscopic coordinates in the case that metrical fluctuations are well below the dynamical scale of the system, i.e. the corresponding energy E is $E \ll E_P$ ($E_P$: Planckian limit). If one builds the expectation value then it follows from Equation (4.1)

$$c_\mu = \langle X'_\mu \rangle - N_\mu^\alpha \langle X_\alpha \rangle \tag{4.3}$$

since $\langle \varsigma_\mu \rangle = 0$. Inserting Equation (4.3) into (4.1) one gets

$$x'_\mu = N_\mu^\alpha x_\alpha + \varsigma_\mu. \tag{4.4}$$

This means, using the assumption in Equation (4.2) it is possible to separate the minimization in Equation (4.1) into a macroscopic relation given by Equation (4.3) and into a microscopic relation given by Equation (4.4). For $c_\mu=0$ Equation (4.3) gives the LT and otherwise one gets Poincaré transformations (PT). The consequences of the separation approach become obvious if Equation (4.4) is inserted into Equation (4.1). Since

$$\langle x_\mu \varsigma_\nu \rangle = 0 \quad \text{but} \quad \langle x'_\mu \varsigma_\nu \rangle \neq 0 \qquad (4.5)$$

and since the components $N_\mu^\sigma$ are c-values due to Equation (4.3) the minimization in Equation (4.1) is fulfilled if

$$\langle x'_\mu x_\beta \rangle = N_\mu^\alpha \langle x_\alpha x_\beta \rangle. \qquad (4.6)$$

Notwithstanding Equation (4.2), one assumes $g_{\mu\nu}=g_{\nu\mu}$. Now, it can be shown that one gets the following solution for Equation (4.6)

$$N_\mu^\beta = \frac{(-1)^{D-1}}{(D-1)!} \det^{-1}(\mathbf{x}) \langle x'_\mu x_\alpha \rangle Q^{\alpha\beta} \qquad (4.7)$$

whereas $Q^{\mu\nu}$ depends on the dimension of the space-time and can be expressed by

$$D=2: \quad Q^{\alpha\beta} = \Delta^{\alpha\sigma} \langle x_\sigma x_\tau \rangle \Delta^{\tau\beta} \qquad (4.8a)$$

$$D=3: \quad Q^{\alpha\beta} = \Delta^{\alpha\sigma\kappa} \langle x_\sigma x_\tau \rangle \langle x_\kappa x_\lambda \rangle \Delta^{\tau\lambda\beta} \qquad (4.8b)$$

$$D=4: \quad Q^{\alpha\beta} = \Delta^{\alpha\sigma\kappa\omega} \langle x_\sigma x_\tau \rangle \langle x_\kappa x_\lambda \rangle \langle x_\omega x_\rho \rangle \Delta^{\tau\lambda\rho\beta} \qquad (4.8c)$$

whereas $\Delta^{\alpha..\beta}$ is the Levi-Civita symbol. The structure of the Equations (4.8a-c) suggests a simple relationship for $\langle x_\sigma, x_\tau \rangle$ given by

$$\langle x_\sigma x_\tau \rangle = s^2 g_{\sigma\tau} \quad \text{and} \quad \langle x_\sigma x^\sigma \rangle = s^2 \text{Tr}(g) \qquad (4.9)$$

whereas $s^2$ considers the length dimension of $x_\sigma$ and is related to an area. Equation (4.9) states that the fluctuations of the microscopic coordinates depend on the metric of the corresponding space. Hence, the Equations (4.2) and (4.9) both reflect the connection between fluctuations and metric. This relationship leads to a comprehensible expression for transformation matrix $\mathbf{N}$ inserting Equation (4.9) into the Equations (4.7) and (4.8)

$$N_\mu^\beta = s^{-2} \langle x'_\mu x_\alpha \rangle g^{\alpha\beta}. \qquad (4.10)$$

An additional constraint occurs by calculating $g'^{\mu\nu}\langle x'_\mu, x'_\nu \rangle$ using the Equations (4.4) and (4.9). One gets

$$s'^2 \text{Tr}(g') = s^2 g'^{\mu\nu} N_\mu^\alpha N_\nu^\beta g_{\alpha\beta} + e^2 \text{Tr}(g') \quad \text{if} \quad \langle \varsigma_\sigma \varsigma_\tau \rangle = e^2 g'_{\sigma\tau}. \qquad (4.11)$$

Making use of Equation (14) it follows

$$e^2 = s'^2 - s^2 > 0. \qquad (4.12)$$

So, the transformation invariance of the inner product required in Equation (14) simplifies the relation that describes the fluctuation induced transformation errors which are attached to $\Sigma'$ for measurements performed in $\Sigma$.

**References**

Balian, Roger (2003) *Entropy, a protean concept*, Poincaré Seminar **2**, pp. 119-145
Collins, John (1995) *The problem of scales: Renormalization and all that*, arXiv: hep-ph/9510276
Jizba, Petr and Arimitsu, Toshihico (2001) *The world according to Rényi: Thermodynamics of fractal systems*, arXiv: cond-mat/0108148
11